# Predictions and correlation analyses of Ellingham diagrams in binary oxides

Shun-Li Shang,[1,2,*] Shuang Lin,[1] Michael C. Gao,[2] Darrell G. Schlom,[3] and Zi-Kui Liu[1]

[1] Department of Materials Science and Engineering, The Pennsylvania State University, University Park, Pennsylvania 16802, USA

[2] Materials Engineering and Manufacturing Directorate, National Energy Technology Laboratory, Albany, Oregon 97321, USA

[3] Department of Materials Science and Engineering, Cornell University, Ithaca, New York 14853, USA

* sus26@psu.edu (S. L. Shang)

NOTE THAT please contact Dr. Shang for the mentioned Excel file if need.

## ABSTRACT:

Knowing oxide-forming ability is vital to gain desired or avoid deleterious oxides formation through tuning oxidizing environment and materials chemistry. Here, we have conducted a comprehensive thermodynamic analysis of 137 binary oxides using the presently predicted Ellingham diagrams. It is found that the "active" elements to form oxides easily are the f-block elements (lanthanides and actinides), elements in the groups II, III, and IV (alkaline earth, Sc, Y, Ti, Zr, and Hf), and Al and Li; while the "noble" elements with their oxides nonstable and easily reduced are coinage metals (Cu, Ag, and especially Au), Pt-group elements, and Hg and Se. Machine learning based sequential feature selection indicates that oxide-forming ability can be represented by electronic structures of pure elements, for example, their d- and s-valence electrons, Mendeleev numbers, and the groups, making the periodic table a useful tool to tailor oxide-forming ability. The other key elemental features to correlate oxide-forming ability are thermochemical properties such as melting points and standard entropy at 298 K of pure elements. It further shows that the present Ellingham diagrams enable qualitatively understanding and even predicting oxides formed in multicomponent materials, such as the Fe-20Cr-20Ni alloy (in wt.%) and the equimolar high entropy alloy of AlCoCrFeNi, which are in accordance with thermodynamic calculations using the CALPHAD approach and experimental observations in the literature.

## 1 Introduction

Knowing oxide-forming ability is critical to examine materials degradation or enhance resistance to materials degradation in oxidation environments [1], and to synthesize desired oxides by tuning growth conditions [2]. For example, it is expected to form and maintain a dense, stable, and continuous scale (e.g., α-$Al_2O_3$, α-$Cr_2O_3$, and $SiO_2$ [1], and $CrTaO_4$ [3]) to provide oxidation and hot corrosion protection of base alloys. Additionally, (multi)functional oxides possess enormous interest for a wide range of applications, hence requiring successful syntheses. These oxides include, for instance, the metallic $PdCoO_2$ [4], the ferroelectric $Pb(Zr,Ti)O_3$ [5], the ferromagnetic $La_{0.7}Sr_{0.3}MnO_3$ [6], the ferrimagnetic $Sr_2FeMoO_6$ [7], the multiferroic $BiFeO_3$ [8] and $BiMnO_3$ [9], the superconducting $HgBa_2Ca_2Cu_3O_{1+x}$ [10], and the topological insulator of $Sr_3SnO$ [11].

Relative stability of desired or deleterious oxides can be examined by the Ellingham diagram, which was introduced by Ellingham in 1944 [12] to plot the change of the standard Gibbs energy ($\Delta G^o$) versus temperature ($T$) for a given reaction; see details in Section 2.1. In the context of oxidation, the Ellingham diagram determines the relative ease by reducing a given oxide to element(s) or other oxide(s), i.e., the lower the line on the Ellingham diagram, the more thermodynamically stable the oxide relative to the other oxides will be; where the "noble" metals (or elements, or oxides) are closer to the top of the diagram, and the "active" ones are closer to the bottom of the diagram. In addition, replacing $\Delta G^o$ with the partial pressure of oxygen (i.e., $P_{O_2}$); see Section 2.1, the Ellingham diagram gives the equilibrium $P_{O_2}$ values for a given reaction, where the element (or oxide) will be oxidized at higher $P_{O_2}$ values or will be reduced at lower $P_{O_2}$ values. The Ellingham diagram is hence a predictive tool to tailor the relative stability of oxides.

For example, the Ellingham diagrams were used to guide the growth of oxide thin films by means of the molecular-beam epitaxy (MBE) such as the adsorption-controlled growths of $BaSnO_3$ [13], $BiFeO_3$ [14], $BiMnO_3$ [9], $LuFe_2O_4$ [15], $PbTiO_3$ [16], $SrRuO_3$ and $CaRuO_3$ [17], and $Sr_3SnO$ [11]. Ellingham diagrams were also used to tailor thermodynamic stability of the spinel $MgAl_2O_4$ under strong reducing conditions [18], the structural phase transition of $Pr_2NiO_{4+\delta}$ involving variation of oxygen content [19], the hot isostatic press processes [20], and oxidation behaviors of transition metals [21], ultrahigh-entropy ceramics [22], and multi-principal component materials [23,24].

Despite the importance, a comprehensive compilation of oxide Ellingham diagrams is still limited in the literature. For instance, Backman and Opila [23] calculated the Ellingham diagrams for only the Groups IV, V, and VI binary oxides using the FactSage software and the Fact Pure Substance database [25]. Birks et al. [26], Ellingham [12], Gleeson [27], and Smeltzer and Young [21] reported only some Ellingham diagrams which are mainly for transitional metal binary oxides.

The present work aims to enrich the literature by providing, first, a compilation of Ellingham diagrams for 137 binary oxides based on a comprehensive thermodynamic analysis. Then, oxide-forming ability as indicated by the Ellingham diagrams is examined through machine learning based correlation analysis in terms of the fundamental properties of pure elements. As applications, thermodynamic stabilities of the Fe-20Cr-20Ni alloy (in wt.%) and the equimolar high entropy alloy AlCoCrFeNi in oxidizing environments are predicted using the CALPHAD (calculations of phase diagram) approach [28] and understood by the presently predicted Ellingham diagrams.

## 2 Methodology

### 2.1 Basis of Ellingham diagram

Ellingham in 1944 [12] introduced the Ellingham diagram to study metallurgic processes involving oxides and sulfides. This diagram shows the plot regarding the change of the standard Gibbs energy ($\Delta G^{o}$) with respect to temperature (*T*) for a given reaction, measuring thermodynamic driving force that makes a reaction occurring. In the present work, the following two scenarios are considered to form binary oxides, where the reactions in "*scenario one*" are from metal/element M to oxide $M_xO_y$,

$$\frac{2x}{y}M(s \text{ or } l) + O_2(g) = \frac{2}{y}M_xO_y(s \text{ or } l) \qquad \text{Eq. 1}$$

and the reactions in "*scenario two*" are from one oxide $M_aO_b$ to another oxide $M_xO_y$,

$$\frac{2x}{ay - bx}M_aO_b(s \text{ or } l) + O_2(g) = \frac{2a}{ay - bx}M_xO_y(s \text{ or } l) \qquad \text{Eq. 2}$$

where the letters *s*, *l*, and *g* indicate the solid, liquid, and gas phases, respectively. Note that (i) ay > bx in Eq. 2, indicating that the oxidation process is from $M_aO_b$ to $M_xO_y$; (ii) the oxygen gas species $O_2$ is selected due to its dominance in comparison with the other species of $O_1$ and $O_3$ [29]; and (iii) the reactions are under the total pressure ($P_{tot}$) of one atmosphere (atm) by consuming one mole of oxygen ($O_2$). Assuming the activities of liquids and solids as unity, the *T*-dependent equilibrium constant $K$ for Eq. 1 and Eq. 2 is determined by the activity of oxygen ($a_{O_2}$) [27],

$$K = 1/a_{O_2} \qquad \text{Eq. 3}$$

For reactions in Eq. 1 and Eq. 2, the Gibbs energy change under isobaric conditions (e.g., at one atm) is given by,

$$\Delta G = \Delta G^o + RT\ln K = \Delta G^o - RT\ln a_{O_2} \quad \text{Eq. 4}$$

where $\Delta G^o$ is the standard Gibbs energy of formation for Eq. 1 and Eq. 2 at absolute temperature $T$, and $R$ is the gas constant. Eq. 4 indicates that if $\Delta G < 0$ the reaction can proceed spontaneously without external inputs, and if $\Delta G > 0$ the reaction is thermodynamically unfavorable. At equilibrium ($\Delta G = 0$), the reaction is at equilibrium and

$$\Delta G^o = RT \ln a_{O_2} \quad \text{Eq. 5}$$

In the present work, we set the reference pressure $P_{ref} = P_{tot} = 1$ atm ($10^5$ Pa), and the unit of $P_{O_2}$ is Torr. If oxygen behaves ideally,

$$a_{O_2} = \frac{P_{O_2}}{P_{ref}} = \frac{133.33\, P_{O_2}}{10^5} \quad \text{Eq. 6}$$

It shows that $P_{O_2} = 750 a_{O_2}$ Torr. Noting that $\Delta G^o = \Delta H^o - T\Delta S^o$, where $\Delta H^o$ and $\Delta S^o$ are the enthalpy change and the entropy change for a given reaction, respectively. Since $\Delta H^o$ and $\Delta S^o$ are essentially constant with respect to temperature, the $\Delta G^o$ vs. $T$ plots are in principle the straight lines with $\Delta S^o$ being the slope (i.e., the molar entropy of oxygen since $\Delta S^o \cong -S^o_{O_2}$ [27] for Eq. 1 and Eq. 2) and $\Delta H^o$ being the intercept.

It is worth mentioning that the $\Delta G^o$ vs. $T$ plot is usually called the Ellingham diagram [12], while the $\ln K$ vs. $1/T$ plot is usually called the Van't Hoff diagram, and both diagrams are identical [30]. In the present work, the "$P_{O_2}$ vs. $1/T$" plot is adopted as the Ellingham *P*-*T* phase diagram to facilitate the determination of $P_{O_2}$ for a given reaction – the same as our previous applications for guiding the MBE growths of oxides [9,11,13–17].

## 2.2 Thermodynamic calculations of Ellingham diagram

Ellingham diagram involves the determination of $\Delta G^o$ or $P_{O_2}$ with respect to $T$ for a given reaction. In the present work, these thermodynamic calculations were performed by the Thermo-Calc software [31] using the SGTE substance database (i.e., the SSUB5) [32]. The missing thermodynamic properties for the Pt-O compounds (PtO, $PtO_2$, and $Pt_3O_4$) were added to SSUB5 using the reported enthalpies and entropies of formation at room temperature [33]. Thermodynamic calculations of $\Delta G^o$ or $P_{O_2}$ can use one of the following three methods.

- Method **I**: Direct calculations of $\Delta G^o$ for a given reaction with $P_{O_2}$ determined by Eq. 5 and Eq. 6. In this case, tabulated $\Delta G^o$ values can be predicted, where only the gas species of $O_2$ is included in the calculations and all the other gas species are excluded.
- Method **II (Eq-reaction):** Equilibrium calculations by considering all phases relevant to the reaction of study. In this case, all gas species (e.g., $O_1$, $O_2$, and $O_3$) are considered and the $P_{O_2}$ values are determined after thermodynamic calculations. In the present work, the Method II is mainly used in terms of a high-throughput way for the calculations.
- Method **III (Eq-system):** Equilibrium calculations by considering all available phases in the system of study, e.g., the Fe-O system, and the Ellingham diagrams for all reactions in in this system can be calculated simultaneously. Method III is in fact the equilibrium calculations of $P$-$T$ phase diagram for a given system; see our previous calculations for ternary oxides of $BaSnO_3$ [13], $BiFeO_3$ [14], $BiMnO_3$ [9], $LuFe_2O_4$ [15], $PbTiO_3$ [16], $SrRuO_3$ and $CaRuO_3$ [17], and $Sr_3SnO$ [11]. However, the nonstable oxides are excluded in Method III and especially the successful calculations are not easy for some systems due partially to the choices of initial conditions.

In principle, all the above three methods predict the same Ellingham diagrams (see Section 3.1 the Fe-O Ellingham diagrams predicted by the Methods II and III). During our calculations of Ellingham diagrams using Method II, the following two situations are considered. First, only the stable or metastable oxides at one atm are included, while the nonstable oxides identified by thermodynamic calculations are excluded (except for the nonstable $Au_2O_3$, which is the only oxide in Au-O based on SSUB5). A total of 137 binary oxides are considered to build the Ellingham diagrams among the 155 oxides available in the SSUB5; see them in Supplemental Excel file (sheet: info-reactions). Second, the reactions of Eq. 1 and Eq. 2 only involve the phases that are able to be in equilibrium with each other, i.e., only the phases involving neighboring oxidation states are allowed in a reaction, making the predictions from both Methods II and III similar. For example, the oxidations in the Cr-O system are from Cr to $Cr_2O_3$ and then to $Cr_5O_{12}$ with increasing oxygen content, where the oxidation from Cr to $Cr_5O_{12}$ is excluded; see a complete list in Supplemental Excel file (sheet: info-reactions).

As an example, one of the Thermo-Calc macro files (i.e., the tcm files) to predict the $Al_2O_3$ Ellingham diagram using Method II is shown in the Supplemental

Table S 1, which includes the phases of α-$Al_2O_3$ (both solid and liquid corundum), δ-$Al_2O_3$, γ-$Al_2O_3$, κ-$Al_2O_3$, Al (both solid and liquid), and all gas species (included but dormant) available in SSUB5, which are all relevant to the reaction of $\frac{4}{3}\mathrm{Al}(s \text{ or } l) + \mathrm{O_2}(g) = \frac{2}{3}\mathrm{Al_2O_3}(s \text{ or } l)$. After equilibrium calculations, the $P_{\mathrm{O_2}}$ value (in Torr) is determined by Eq. 6; see also the function in the Supplemental tcm file.

# 3 Results and discussion

## 3.1 Ellingham diagrams of binary oxides

As an example, Figure 1 shows the predicted Fe-O Ellingham diagram using both the Method II (Eq-reaction) and the Method III (Eq-system), *cf.*, Section 2.2. Figure 1a (by Method III) clearly identifies the predicted phase regions including the gas phase region, while Figure 1b (by Method III) shows in principle the individual lines for the reactions of study, but the associated phase regions are identical to those in Figure 1a. The red-dashed lines in Figure 1b are for reference only, since the nonstable $Fe_{0.947}O$ was ignored in the present work. The bule-solid lines in both Figure 1a and Figure 1b are identical, indicating that both the Methods II and III predict the same results. In addition, the black dots in both diagrams indicate the three-phase equilibria due to such as the solid/solid, solid/liquid, or liquid/gas phase transition. For example, there are two black dots for the reaction of $2\mathrm{Fe} + \mathrm{O_2} = 2\mathrm{FeO}$, representing the three-phase equilibria of "FeO (*l*), Fe (*l*), and Fe (*s*3)" and "FeO (*l*), FeO (*s*), and Fe (*s*3)", with the (T, $P_{\mathrm{O_2}}$) values at (1811 K, $4.06\times10^{-7}$ Torr) and (1650 K, $1.94\times10^{-8}$ Torr), respectively. A complete list of the possible three-phase equilibria for all the present reactions is given in the Supplemental Excel file (sheet: info-reactions).

Figure 2 summarizes the predicted Ellingham diagrams for 137 reactions to form binary oxides with the data of these plots given in the Supplemental Excel file (sheet: TP-data), where 74 of them belong to the reactions in *scenario one* (*cf.,* Eq. 1) and 63 in *scenario two* (*cf.,* Eq. 2). On the top of Figure 2 with higher $P_{O_2}$ values, the metals (elements) are hard to be oxidized; while at the bottom of Figure 2 with lower $P_{O_2}$ values, the metals (elements) are easily to be oxidized. The $P_{O_2}$ values at a given temperature hence can rank thermodynamic stability, i.e., the oxide-forming ability, of the oxides. The lower the $P_{O_2}$ value, the higher the oxide-forming ability will be, and *vice versa*. It shows that the most stable binary oxide is $Tb_2O_3$ associated with the reaction of $\frac{4}{3}\mathrm{Tb} + \mathrm{O_2} = \frac{2}{3}\mathrm{Tb_2O_3}$, which is the lowest line in the diagram, implying that the *f*-element Tb can reduce all the oxides of all other metals. The most nonstable oxide is $Au_2O_3$, indicating that the reaction $\frac{4}{3}\mathrm{Au} + \mathrm{O_2} = \frac{2}{3}\mathrm{Au_2O_3}$ to form $Au_2O_3$ on the top of Figure 2 is likely impossible or it occurs at extremely high $P_{O_2}$ values (> $6\times10^{10}$ Torr). The commonly observed oxides $Cr_2O_3$ in stainless steel and especially $Al_2O_3$ in nickel-based superalloy are easily to be formed due to the lower $P_{O_2}$ values required; see Figure 2.

To facilitate the analysis, Figure 3 lists the $\log P_{O_2}$ values at a selected temperature of 1100 K for the 137 reactions, see their values in Supplemental Excel file (sheet: info-reactions); indicating that the highest $\log P_{O_2}$ (= 11.86) is for the formation of $Au_2O_3$ and the lowest $\log P_{O_2}$ (= -49.21) is for $Tb_2O_3$. In general, the $\log P_{O_2}$ (or $P_{O_2}$) values increase with increasing oxygen content for a given system at a given temperature. For example, the $\log P_{O_2}$ values in the Ti-O system increase from -38.57 to -21.36 at 1100 K with increasing oxygen content from the formation of TiO to $Ti_2O_3$, $Ti_3O_5$, $Ti_4O_7$, and up to $TiO_2$; see also the Supplemental Excel file. However, at a given temperature such as 1100 K, the $\log P_{O_2}$ values do not already increase with increasing oxygen content when metastable oxides exist. These cases include the metals of K, Mn, Mo, Rh, Sn, Cs, Re, and Pu. For example, the $\log P_{O_2}$ values are -17.34, 3.49, and -0.88 at 1100 K for the reactions of $4K + O_2 = 2K_2O$, $2K_2O + O_2 = 2K_2O_2$, and $K_2O_2 + O_2 = 2KO_2$, respectively; see also the Supplemental Excel file (sheet: info-reactions) as well as the calculated K-O pressure-temperature phase diagram in Supplemental Figure S 1.

To illustrate the trends, Table 1 shows and colors the $\log P_{O_2}$ values at 1100 K in the periodic table for the 74 *scenario one* reactions, *cf.,* Eq. 1. The "active" elements to form oxides easily are colored by green and with lower $\log P_{O_2}$ values, including the f-block and group IIIB elements (lanthanides, actinides, Sc, and Y) and their neighbors in group IIA (alkaline earth metals) and group IVB (Ti, Zr, and Hf), and elements Al and Li. These active elements could form very stable oxides such as $Tb_2O_3$, $Y_2O_3$, $Sc_2O_3$, $Er_2O_3$, $Ho_2O_3$, and CaO; see Figure 3. Table 1 indicates that the "noble" elements are colored by red with higher $\log P_{O_2}$ values to form oxides, including the coinage metals (Cu, Ag, and especially Au), Pt-group elements (Ru, Rh, Pd, Os, Ir, and Pt), and Hg and Se. These noble elements could form nonstable oxides and they are easily to be reduced such as $Au_2O_3$, $Ag_2O$, HgO, PtO, PdO, $SeO_2$, and $IrO_2$; see Figure 3.

## 3.2 Understanding oxide-forming ability by correlation analysis

Oxide-forming ability (represented by, for example, the $\log P_{O_2}$ values as shown in Table 1) can be understood and analyzed by a set of elemental attributes/features in terms of the sequential feature selection (SFS) method using a suitable machine learning (ML) algorithm [34]. On the basis of our previous work [34] and the present examinations, the selected ML algorithm is the Gaussian Process Regression (GPR) with the kernel function of matern52 [35] as implemented in the software MATLAB (version R2021b). The presently examined features include the quantities of pure elements related to

- Periodic table: e.g., atomic number, period, group, and Mendeleev number (M_Num2),
- Electronic configurations: the filled and unfilled s, p, d, f, and total valence electrons,
- Atomic size: e.g., covalent radius, van der Waals atomic radius, and atomic volume,
- Physical properties: e.g., thermal conductivity, electrical conductivity, electron density, electron affinity, and electronegativity (EleNeg),
- Thermochemical properties: e.g., heat of sublimation, heat of vaporization (VaporHeat), heat capacity, standard entropy at 298 K (S298), melting temperature (MeltingT), and boiling temperature, and
- Elastic properties: bulk modulus, shear modulus, and Young's modulus.

All the 42 elemental features are explained in Supplemental Table S 2 with their values given in Supplemental Excel file (sheet: 42_features).

After a few iterations using MATLAB between the SFS processes and the GPR verifications by means of the 5-fold cross-validation (CV) for at least 500 times (see details in [34]), the ranked features based on their statistical significances are listed in Supplemental Table S 2 and Figure S 2, showing the overall mean absolute error (MAE) values of $\log P_{O_2}$ (for reactions in *scenario one*) with respect to the combined features starting from the single feature of NdVal (the number of d valence electrons). Here the feature NdVal has the highest correlation with respect to $\log P_{O_2}$ based on the $R^2$ values of linear fit-of-goodness; see Supplemental Table S2. Figure 4 shows that these exists a strong correlation between the d valence electrons (NdVal) and $\log P_{O_2}$ with $R^2$ = 0.73 and MAE = 7.2. Adding more features in addition to NdVal results in the increase of $R^2$ and the decrease of MAE. Table 2 lists the optimal set of 15 features, reaching an overall MAE = 2.00 (Figure S 2) and $R^2$ = 0.98 (not shown). Especially, the top 6 features, which decrease the MAE faster (*cf.*, Figure S 2 and Table 2), are NdVal, NsVal, M_Num2, Group, MeltingT, and S298. It concludes that the trends observed in the periodic table (Table 1) are in fact correlated with the electronic structures of pure elements such as the features of NdVal, NsVal, M_Num2, and Group. In addition, MeltingT and S298 are also the key features to tailor oxide-forming ability, besides minor effects of the other thermochemical and physical properties such as the heat of vaporization, heat capacity, and electronegativity; see Figure S 2 and Table 2.

As one of the examinations concerning the optimal set of 15 features, Figure 5 shows the predicted $\log P_{O_2}$ values (for reactions in *scenario one* at 1100 K) by GPR using the 5-fold CV, illustrating a good agreement with the original $\log P_{O_2}$ values with the fitted $R^2$ = 0.976 ± 0.006. The outliers are the formations of $Cr_2O_3$, HgO, BeO, and $Ag_2O$ based on at least 2000 machine learning trainings; see detailed methodology in [34]. Figure 5 indicates the statistical significances of the present 15 features (Table 2), which correlate oxide-forming ability of pure elements.

The present correlation analysis demonstrates that (i) electronic structures of pure elements are the key to regulate their oxide-forming ability, resulting in obvious trends regarding the $\log P_{O_2}$ values in the periodic table in Table 1; and (ii) the thermochemical properties such as MeltingT and S298 are secondly key features to regulate oxide-forming ability.

### 3.3 Examining oxide-forming ability in multicomponent alloys

Ellingham diagrams of binary oxides are fundamental to understand and even predict oxide-forming ability in multicomponent alloys. Taking concentrated alloys as example, Figure 6 shows the predicted phases as a function of $P_{O_2}$ for the equimolar high entropy alloy AlCoCrFeNi at 1000 K and the Fe-20Cr-20Ni (wt.%) alloy at 873 K, where the CALPHAD results were performed by the Thermo-Calc software and the TCFE8 database [31]. Figure 6a shows that with increasing $P_{O_2}$ from such as $10^{-40}$ Torr (i.e., $\log P_{O_2}$ = -40), first the $Al_2O_3$-rich corundum forms and then the $Cr_2O_3$-rich corundum appears ($P_{O_2} > 10^{-27}$ Torr or $\log P_{O_2} > -27$) in AlCoCrFeNi, while the other oxides such as FeO and NiO are not observed. The CALPHAD-based calculations agree with experimental observations where the formed oxides in AlCoCrFeNi are mainly $Al_2O_3$ and $Cr_2O_3$ after 25 h oxidation at 1000 K [24]. Table 1 shows that $\log P_{O_2}$= -39.2 and -24.1, respectively, for the formations of $Al_2O_3$ and $Cr_2O_3$ at 1100 K, agreeing well with experimental observations and the present CALPHAD predictions. The unobserved oxides are due mainly to the higher $P_{O_2}$ values required to form, e.g., $\log P_{O_2}$ = -10.5 for NiO, $\log P_{O_2}$ = -12.0 for CoO, and $\log P_{O_2}$ = -16.5 for FeO at 1100 K; *cf.*, Table 1 as well as the Supplemental Excel file.

Figure 6b shows that with increasing $P_{O_2}$ from such as $\log P_{O_2}$ = -31, first the $Cr_2O_3$-rich corundum appears in Fe-20Cr-20Ni at 873 K, then the Fe-Cr rich and the Fe-rich spinel appears when $\log P_{O_2}$ > $10^{-28}$ and $10^{-22}$, respectively. Experimentally the observed oxides are mainly $Cr_2O_3$ at 873 K after 1000 h oxidation in air [36]. The unobserved NiO in both CALPHAD predictions and experiments is due mainly to the high $P_{O_2}$ value to form NiO, while the unobserved FeO is due mainly to the relatively high $P_{O_2}$ value required; for example, log$P_{O_2}$ = -24.1 for $Cr_2O_3$, log$P_{O_2}$ = -16.5 for FeO, log$P_{O_2}$ = -10.5 for NiO at 1100 K as shown in Table 1 as well as the Supplemental Excel file.

## 4 Summary and concluding remarks

The present work provides first a calibration of Ellingham diagrams for 137 binary oxides based on a comprehensive thermodynamic study, showing the "active" elements to form oxides easily are the f-block elements (lanthanides and actinides), the elements in groups II, III, and IV (alkaline earth, Sc, Y, and Ti, Zr, and Hf), and the elements Al and Li; and the "noble" elements associated with nonstable oxides are coinage metals (Cu, Ag, and especially Au), Pt-group elements, and Hg and Se. Then, correlation analysis via the sequential feature selection method has been used to examine the forming ability of binary oxides with respect to the elemental attributes, indicating that electronic structures of pure elements are the key features to tailor oxide-forming ability such as the d- and s-valence electrons, Mendeleev number, and the groups of pure elements. These features make the periodic table a great tool to show oxide-forming ability of pure elements. In addition, thermochemical properties such as melting points and standard entropy at 298 K are also key features to regulate oxide-forming ability. As applications of the present Ellingham diagrams, thermodynamic stabilities of the Fe-20Cr-20Ni alloy (in wt.%) and the equimolar AlCoCrFeNi in oxidation environments are predicted using the CALPHAD approach. The predicted oxides are in accordance with experimental observations, which can be understood by the presently predicted Ellingham diagrams.

It should be remarked that the Ellingham diagram could qualitatively suggest whether a given oxide system will oxidize or reduce at elevated temperatures. To facilitate the comparison, the Ellingham diagram is built under the conditions of total pressure of one atmosphere by consuming one mole of oxygen ($O_2$). However, these conditions may be difficult from the growing conditions of oxides by using such as the MBE [29], which are usually synthesized at a fixed partial pressure of oxygen rather than a fixed total pressure, and the MBE process behaves like either an open or a closed system depending on the fluxes of metallic elements, see the discussion regarding the thermodynamics of MBE (TOMBE) diagram [17,29]. In addition, the Ellingham diagram does not quantify the rate of reaction as well as the effect of elemental diffusivity in both alloys and oxides, which need to be considered for a complete analysis [23,27]. Solid solubilities may exist in oxides, and complex ternary and higher oxides may be formed, which were excluded in the present predictions. In addition, internal oxidation in the subsurface may also need be considered to study oxidation and hot corrosion protection of alloys [27,37].

## Acknowledgements

This work was supported by the U.S. Department of Energy (DOE) through Grant No. DE-AR0001435. SLS also would like to acknowledge the support by the DOE High Performance Computing for Energy Innovation (HPC4EI) on the project "Accelerating High Temperature Operation Development of High Entropy Alloys via High Performance Computation" and through an appointment to DOE Faculty Research Program at the National Energy Technology Laboratory (NETL) administered by the Oak Ridge Institute for Science and Education.

**Tables and Table Captions**

Table 1. Predicted $\log P_{O_2}$ values at 1100 K for the reactions in *scenario one* to form binary oxides, where the green – yellow – red color scale is used to mark the values from low to high.

Reactions in *scenario one* to form binary oxides

| | | | | | | | | | | | | | | | |
|---|---|---|---|---|---|---|---|---|---|---|---|---|---|---|---|
| Li -40.0 | Be -44.7 | | | | | | | | | | | B -28.9 | C | N | O |
| Na -22.5 | Mg -42.9 | | | | | | | | | | | Al -39.2 | Si -31.0 | P -16.2 | S |
| K -17.3 | Ca -46.5 | Sc -47.3 | Ti -38.6 | V -28.2 | Cr -24.1 | Mn -26.0 | Fe -16.5 | Co -12.0 | Ni -10.5 | Cu -5.5 | Zn -19.6 | Ga -20.3 | Ge -14.6 | As -10.0 | Se 1.0 |
| Rb -16.3 | Sr -42.8 | Y -47.3 | Zr -39.4 | Nb -27.5 | Mo -15.7 | Tc -8.4 | Ru -2.9 | Rh -1.4 | Pd 2.7 | Ag 6.4 | Cd -11.1 | In -15.1 | Sn -13.6 | Sb -10.7 | Te -3.0 |
| Cs -17.0 | Ba -39.2 | | Hf -40.6 | Ta -26.9 | W -15.7 | Re -8.4 | Os -1.7 | Ir 0.6 | Pt 5.3 | Au 11.9 | Hg 5.9 | Tl -5.0 | Pb -7.5 | Bi -5.5 | Po |
| | La -44.0 | Ce -44.0 | Pr -44.5 | Nd -44.6 | Pm -44.5 | Sm -44.7 | Eu -42.9 | Gd -45.4 | Tb -49.2 | Dy -45.8 | Ho -46.7 | Er -47.1 | Tm -46.4 | Yb -44.7 | Lu -46.4 |
| | Ac | Th -45.7 | Pa | U -39.6 | Np -39.0 | Pu -40.9 | Am -42.2 | | | | | | | | |

Table 2. The selected 15 elemental features to correlate oxide-forming ability (i.e., the $\log P_{O_2}$ value) based on the sequential feature selection method using the Gaussian process regression algorithm, where the top six features are highlighted by bold and italic texts and the explanations of these features are given in the Supplemental Table S 2.

| Category | Selected elemental features |
|---|---|
| Electronic configurations | ***NdVal (number of d valence electrons), NsVal,*** NfUnfill (number of unfilled f valence electrons), NdUnfill |
| Positions in the periodic table | ***M_Num2 (Mendeleev number), Group,*** Period |
| Thermochemical properties | ***MeltingT (melting temperature), S298 (standard entropy at 298 K),*** VaporHeat (vaporization heat), CohEnergy, Heat_Capacity, BoilingT |
| Physical properties | EleNeg_Miedema, Ion_Pot_3 |

**Figures and Figure Captions**

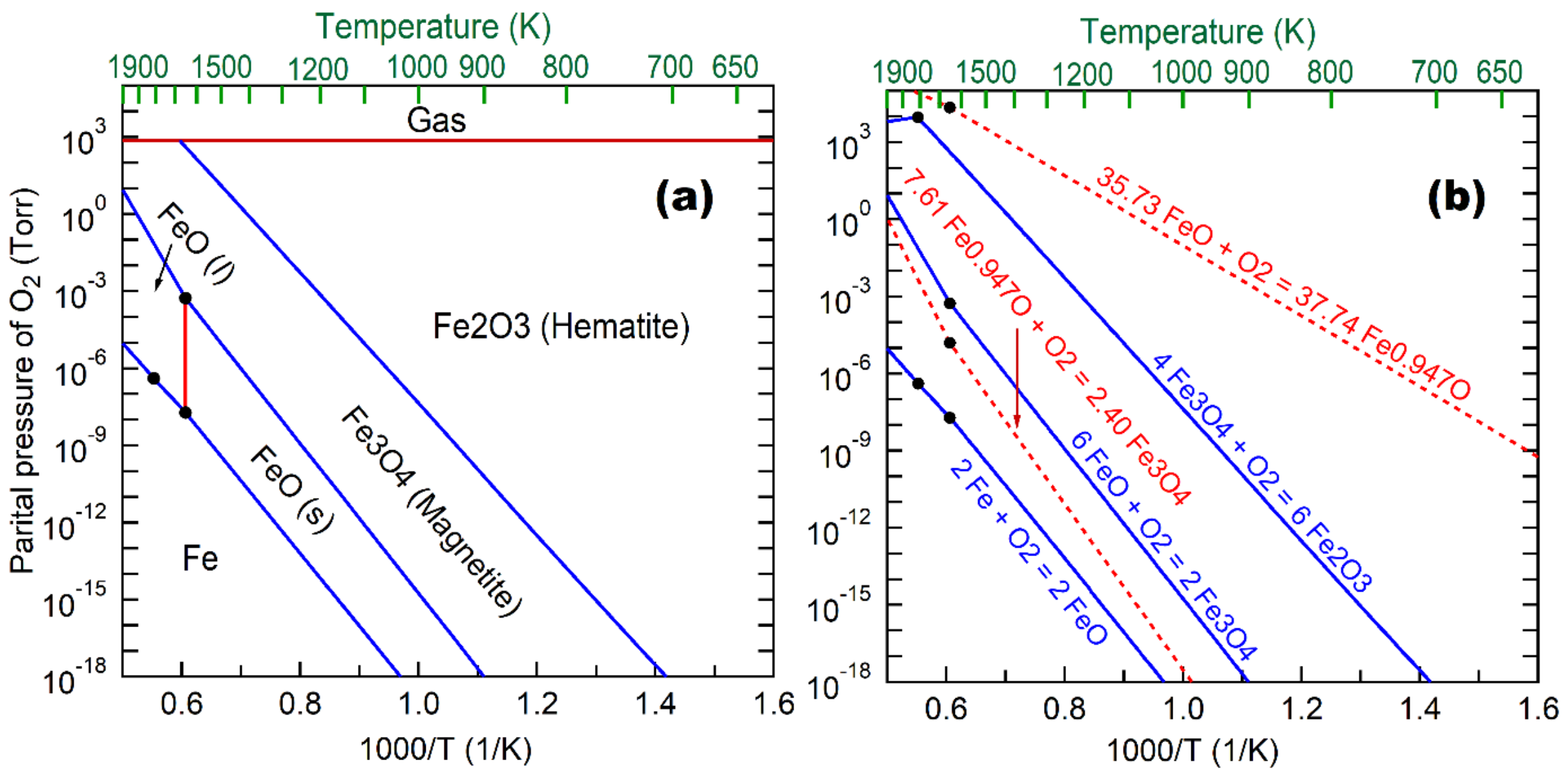

Figure 1. Fe-O Ellingham diagram predicted by (a) the Method II (Eq-reaction) and (b) the Method III (Eq-system). The black dots indicate the solid/solid, solid/liquid, or liquid/gas phase transition. The blue lines in both diagrams are identical, and the red-dashed lines indicate the reactions to form nonstable oxides (ignored in the present work).

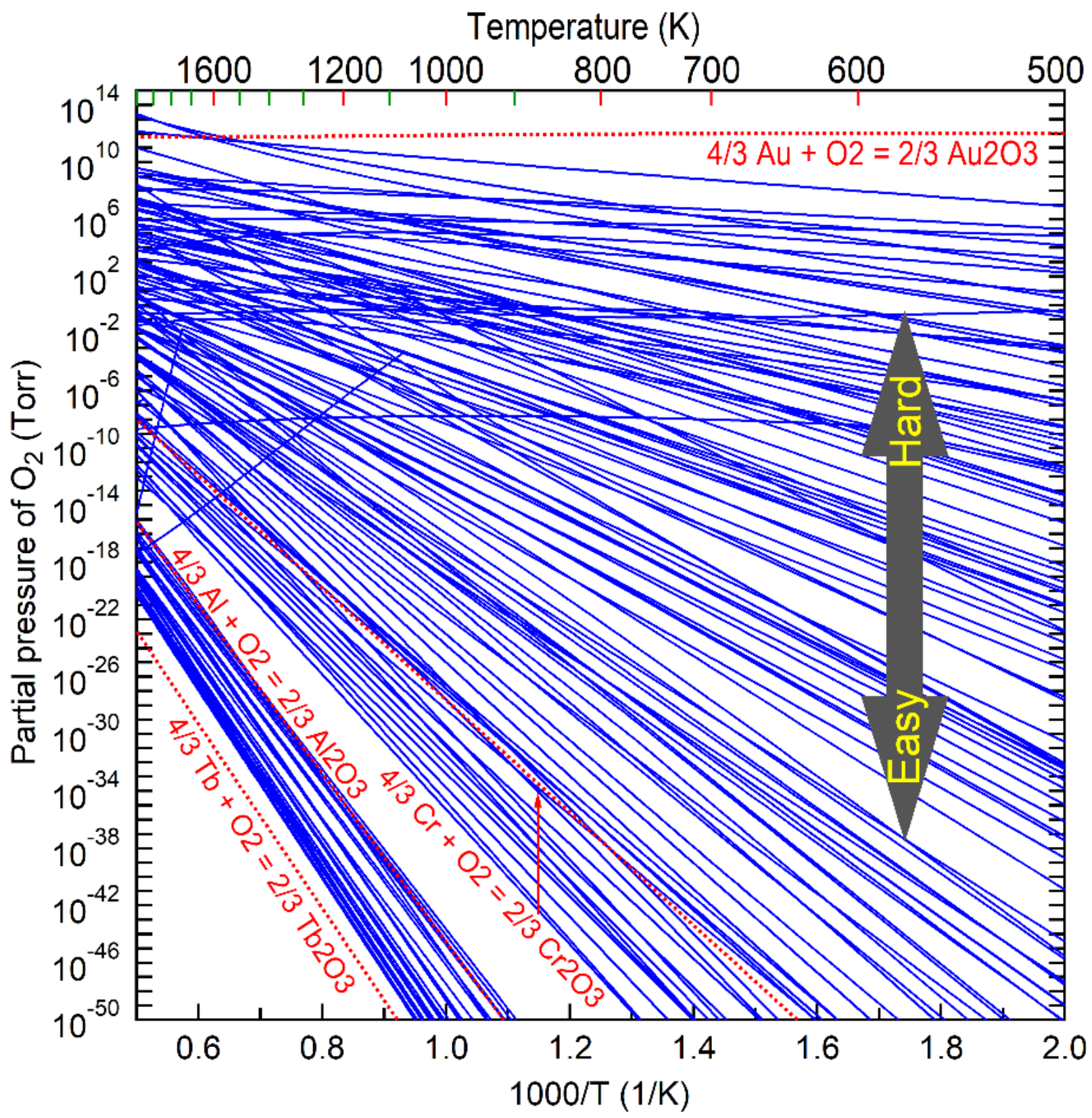

Figure 2. Overview of the predicted Ellingham diagrams to form 137 binary oxides. All data to plot this diagram are shown in the Supplemental Excel file (sheet: TP-data).

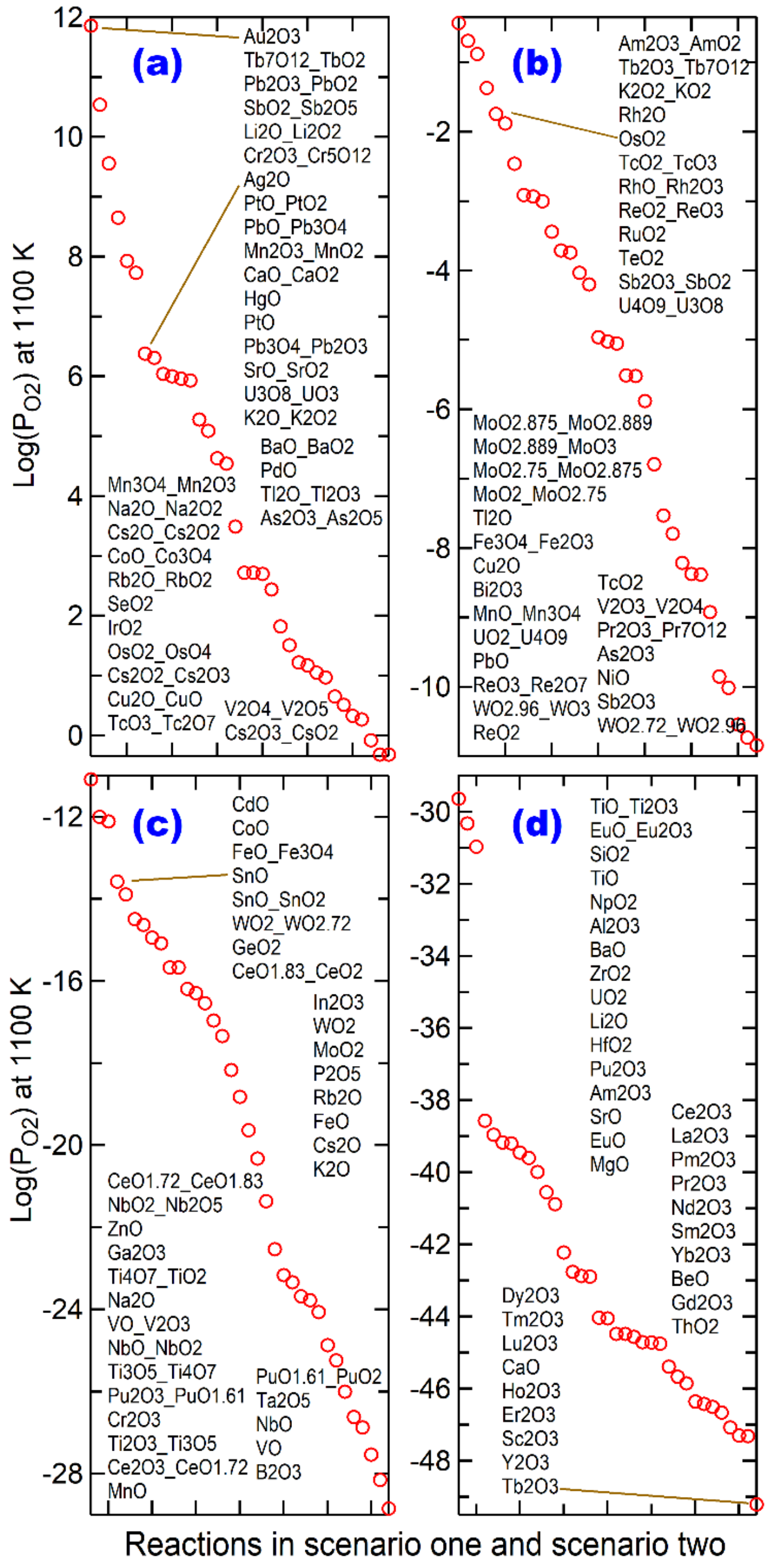

Figure 3. Log$P_{O_2}$ values at 1100 K for the reactions in *scenario one* (74 reactions indicated by one oxide) and *scenario two* (63 reactions indicated by two oxides separated by "_"). The log$P_{O_2}$ values from high to low are shown in the figures (a), (b), (c), to (d).

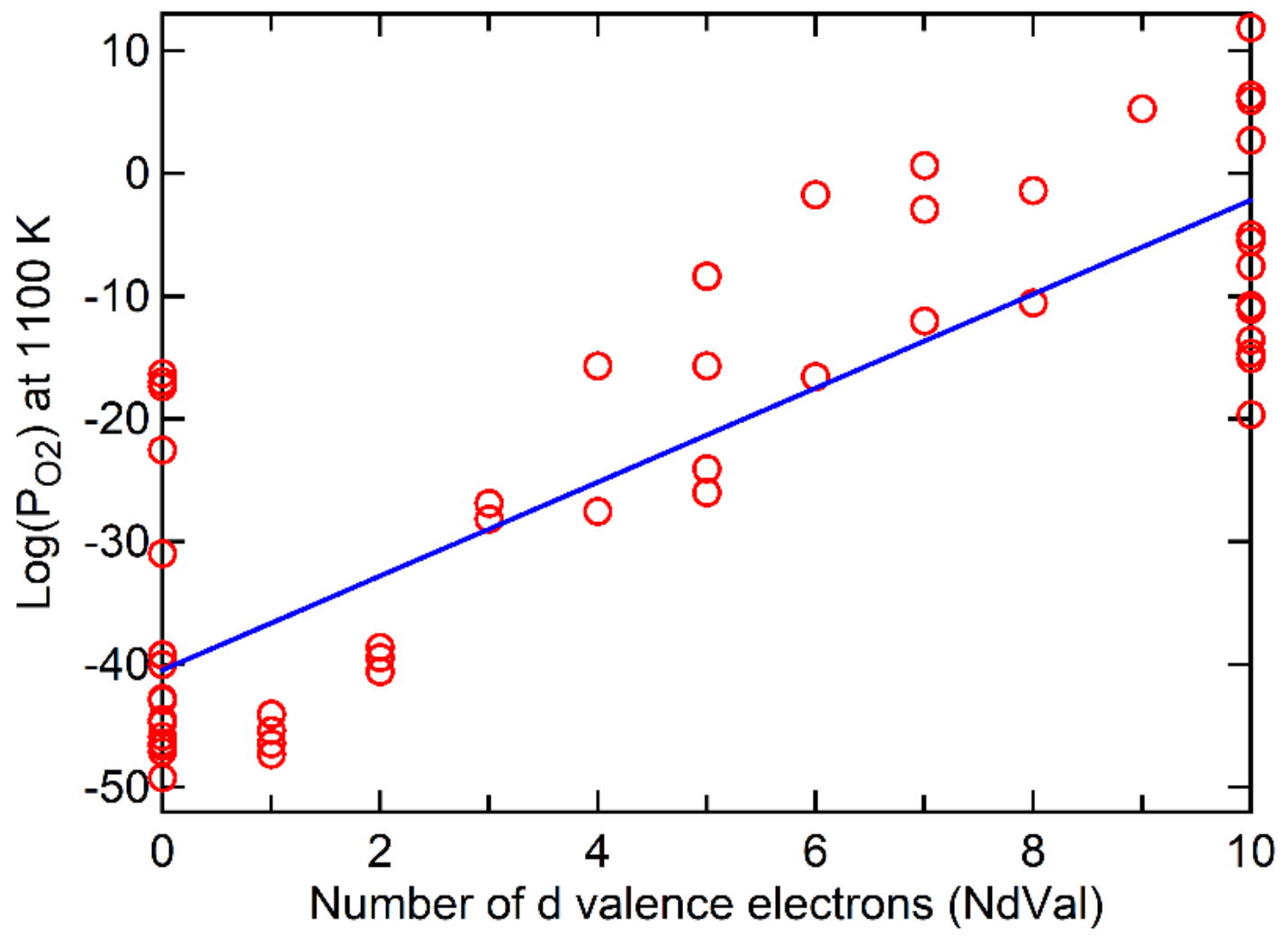

Figure 4. Predicted $\log P_{O_2}$ values for the reactions in *scenario one* (*cf.,* Eq. 1) with respect to the numbers of d valence electrons (NdVal) of pure elements, where the solid line is a linear fit with $R^2 = 0.73$.

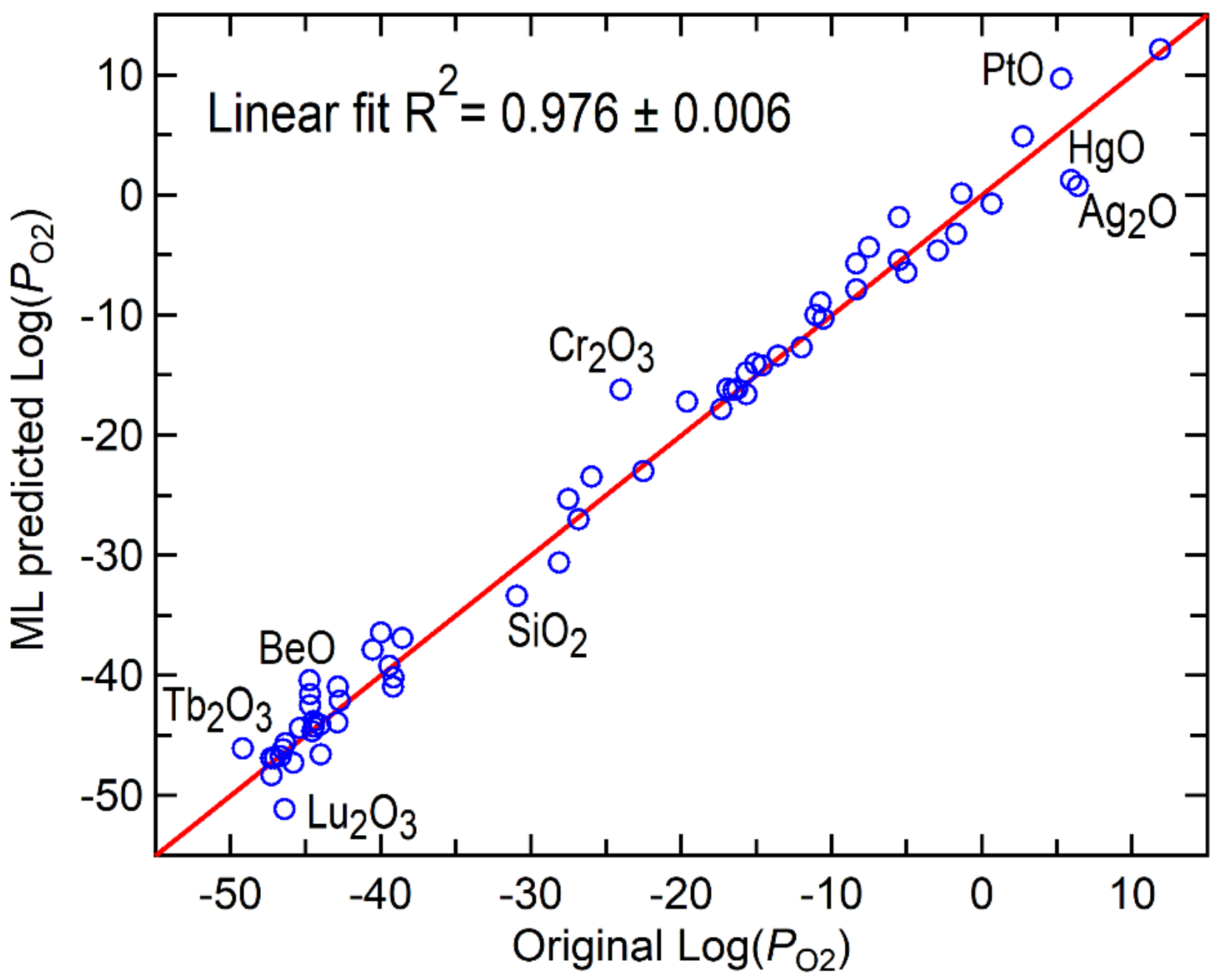

Figure 5. Machine learning (ML) fitting of the log$P_{O_2}$ values using the Gaussian process regression with the kernel function of matern52, 15 features, and the 5-fold CV. This figure shows only one of the fittings. The outliers are the formations of $Cr_2O_3$, HgO, BeO, and $Ag_2O$ and the overall $R^2$ value is based on 2000 ML trainings.

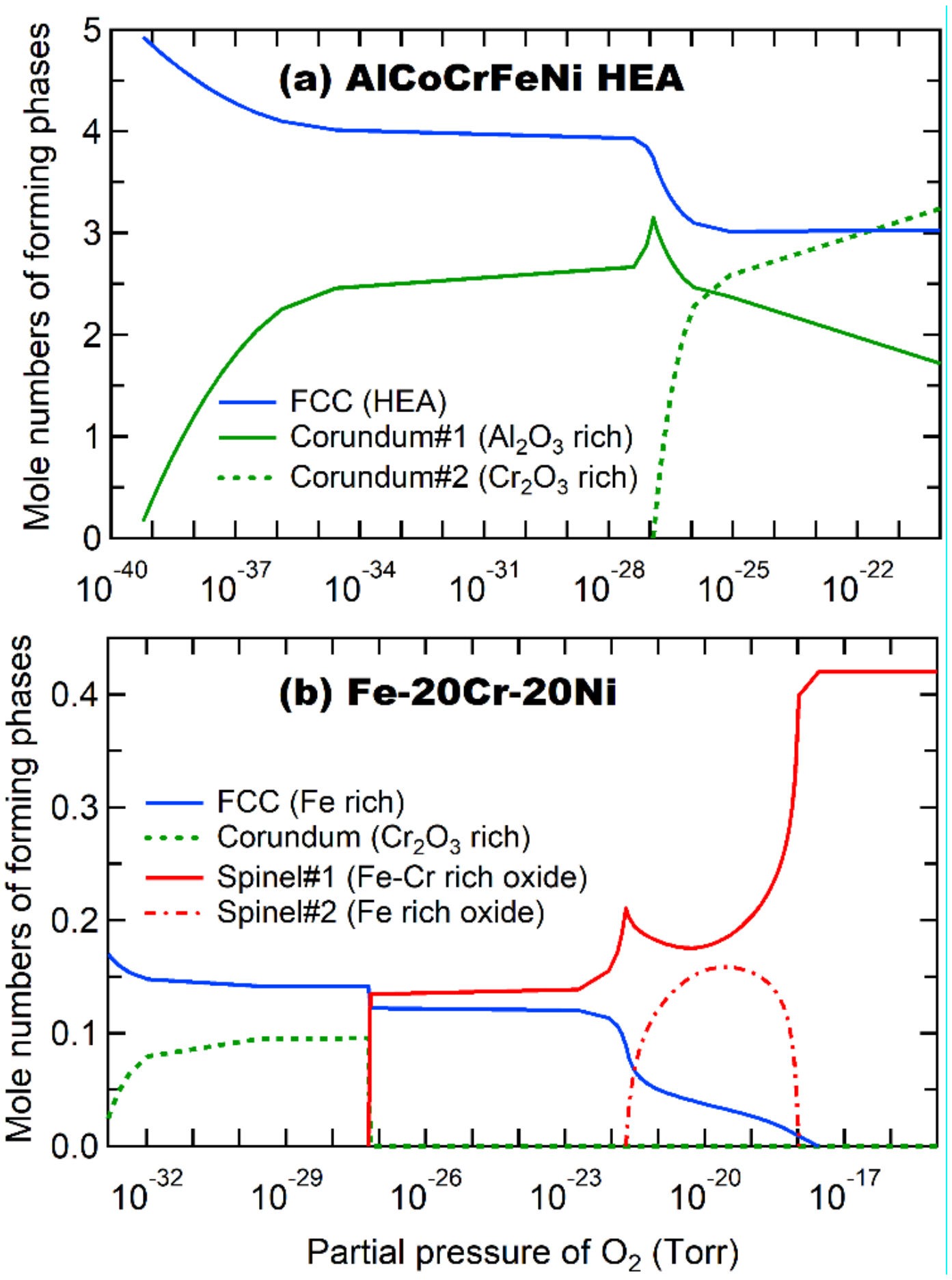

Figure 6. Predicted phases of two alloys as a function of partial pressure of $O_2$ ($P_{O2}$) based on the TCFE8 database: (a) Equimolar AlCoCrFeNi at 1000 K and (b) Fe-20Cr-20Ni (wt.%) at 873 K. Experimentally, the formed oxides in AlCoCrFeNi after 25 h oxidation at 1000 K are mainly $Al_2O_3$ and $Cr_2O_3$ [24], and the formed oxide in Fe-20Cr-20Ni after 1000 h oxidation in air and at 873 K is mainly $Cr_2O_3$ [36].

## Supplementary Material

The **Supplemental** Excel file include three sheets:

- (1) Sheet "info-reactions":
  - Columns A and B list atomic numbers and names of pure elements.
  - Columns C and D list the possible oxides in the SSUB5 database as well as their names when available.
  - Column E (O-vs-M) lists the ratio of oxygen vs. metal (element) for each oxide.
  - Column F (log10_PO2_1100) lists the log (base 10) values of $P_{O_2}$ at 1100 K. Note that for each M-O system, if oxygen content (column E: "O-vs-M") increases, but the values of "Log10_PO2_1100" do not increase, indicting some oxides in this system are metastable; see such as the P-T phase diagram of the K-O system.
  - Column G (Stable-or-not): After examining all phase diagrams for all the M-O systems by thermodynamic calculations, this column shows all the stable and metastable oxides (marked by "Yes") and the nonstable oxides (marked by "No").
  - Column I (Reaction) lists the reactions which are in *scenario one* (showing only one oxide) or in *scenario two* (showing two oxides separated by "_").
  - Column J (No_Reaction) lists the number of each reaction which is for reference only.
  - Column K (Three-Phases-and-T-P) lists all the three-phase equilibrium points and the corresponding temperature and $P_{O_2}$ values.
- (2) Sheet "TP-data":

- Note that (I) the column names starting with "x" indicate the inverse temperatures (1000/T) and the column names starting with "y" indicate the $P_{O_2}$ values (in Torr), where the reactions in *scenario one* are marked by only one oxide and the reactions in *scenario two* are marked by two oxides separated by "_".
- Note that (II) the names of these oxides in this sheet are the same as those in the SSUB5 database.
- For example, columns A and B, which list the inverse temperatures (1000/T, marked by xAG2O1) and the $P_{O_2}$ values (in Torr, marked by yAG2O1) for the reaction in *scenario one* to form $Ag_2O$.

- (3) Sheet "42-features":
  - List values of the 42 elemental features examined in the present work together with the corresponding $P_{O_2}$ values at 1100 K (column AR: logpo2), see the explanations of these features in Table S 2.

Table S 1. An example of the Thermo-Calc macro file (i.e., the tcm file) to predict the $Al_2O_3$ Ellingham diagram using the Method II.

```
go data
sw ssub5
def-ele AL O
get
go p-3
ch ph *=dor
ch ph AL_S AL_L  = e 1
ch ph CORUNDUM CORUNDUM_L AL2O3_DELTA AL2O3_GAMMA AL2O3_KAPPA  = e 1
s-c n(O)=2
s-c n(AL)=1.333333333
s-c p=1e5 t=2000
c-e
c-e
s-a-v 1 t 300 3000,,,,
step,,,,
post
e-sy f torr=acr(o2,gas)*p/133.33;
s-d-a x t-k
s-d-a y torr
s-a-ty y log
s-a-ty x inv
s-a-te x n 1/T
s-a-te y n PO2 (Torr)
s-la b
pl,,,,,,,
set_inter
```

Table S 2. Elemental features/descriptors examined in the present work together with their goodness-of-fit $R^2$ values between the features and the $\log P_{O2}$ values at 1100 K for 74 reactions in *scenario one*. The numbers in "Rank" column rank statistical significances of the features to regulate $\log P_{O2}$ after iterations between the sequential feature selection and the predictions using the Gaussian process regression (with the kernel function of matern52) [34]. Note that the values of these features are given in the Supplemental Excel file (sheet: 42-features).

| Rank | Features | $R^2$ | Explanations |
|---|---|---|---|
| 1 | NdVal | 0.727 | Number of d valence electrons. |
| 2 | M_Num2 | 0.408 | Mendeleev number (MN2, start from left bottom, down-top sequence) [38]. |
| 3 | Group | 0.065 | Group of pure elements in the periodic table. |
| 4 | MeltingT | 0.004 | Melting temperature (in K) based on the collections by Kittel [39]. |
| 5 | S298 | 0.073 | Standard entropy at 298 K (in J/mol.K) [40] |
| 6 | NsVal | 0.204 | Number of s valence electrons. |
| 7 | NfUnfill | 0.201 | Number of unfilled f valence electrons. |
| 8 | Period | 0.002 | Group of pure elements in the periodic table. |
| 9 | VaporHeat | 0.018 | Vaporization heat (in kJ/mol) based on the collections of Wolfram Mathematica (WM); see "VaporizationHeat" in WM [41]. |
| 10 | CohEnergy | 0.030 | Cohesive energy (in eV/atom) collected by Kittel [39]. |
| 11 | EleNeg_Miedema | 0.442 | Electronegativity (in Volt) used in the Miedema model [42]. |
| 12 | Heat_Capacity | 0.042 | Heat capacity at 298 K (in J/kg-mol.K) [43]. |
| 13 | BoilingT | 0.018 | Boiling temperature (in K) [43]. |
| 14 | Ion_Pot_3 | 0.004 | The third ionization potentials to remove two electrons (in eV) [43]. |
| 15 | NdUnfill | 0.044 | Number of unfilled d valence electrons. |
| 16 | NpVal | 0.053 | Number of f valence electrons. |
| 17 | Therm_Conduc | 0.178 | Thermal conductivity at 300 K (in W $cm^{-1}$ $K^{-1}$) [39,44]. |
| 18 | WorkFunc | 0.430 | Work function of pure elements [45]. |
| 19 | Heat_Sublimation | 0.076 | Heat of sublimation at 298 K (in J/mol) [43]. |
| 20 | Ion_Pot_2 | 0.038 | The second ionization potentials to remove two electrons (in eV) [43]. |
| 21 | Number | 0.032 | Atomic number of pure elements in the periodic table. |
| 22 | Ele_Conduc | 0.079 | Electrical conductivity of metals in $(ohm\text{-}cm)^{-1}$ [39]. |
| 23 | NUnfill | 0.156 | Number of total unfilled valence electrons. |
| 24 | Heat_Fusion | 0.015 | Heat of fusion at 298 K (in J/mol) [43]. |
| 25 | V0_Miedema | 0.047 | Atomic volume (in $cm^3$/mol) used in the Miedema model [42] |
| 26 | Radius_Coval | 0.282 | Covalent radius (in pm) based on the collections of Wolfram Mathematica (WM); see "ElementData" in WM [41]. |
| 27 | NsUnfill | 0.155 | Number of unfilled s valence electrons. |
| 28 | DebyeT | 0.002 | Debye temperature (in K) collected by Kittel [39]. |
| 29 | Mass | 0.030 | Mass of pure elements. |
| 30 | Radius_vDW | 0.094 | Van der Waals atomic radius (in pm) [41,46]. |
| 31 | EleDensity_Miedema | 0.219 | Electron density at the boundary of Wigner-Seitz cell used in the Miedema model [42]. |
| 32 | G_wiki | 0.152 | Shear modulus (in GPa) of pure elements based on Wikipedia and Azom [47,48]. |
| 33 | Y_wiki | 0.165 | Young's modulus (in GPa) of pure elements based on Wikipedia and Azom [47,48]. |
| 34 | No_Spectral_lines | 0.017 | Number of spectral lines of the elements [43]. |
| 35 | NpUnfill | 0.037 | Number of unfilled p valence electrons. |
| 36 | EleNeg_Pauling | 0.593 | Electronegativity by Pauling scale (a dimensionless quantity) [41,46]. |
| 37 | NfVal | 0.009 | Number of f valence electrons. |
| 38 | Ion_Pot_1 | 0.308 | The first ionization potentials to remove one electron (in eV) [43] |
| 39 | Electron_Affinity | 0.258 | Electron affinity (in eV) [46]. |
| 40 | B_wiki | 0.275 | Bulk modulus (in GPa) of pure elements based on Wikipedia and Azom [47,48]. Note that elastic properties of fcc Sr were taken from [49]. |
| 41 | Nval | 0.264 | Number of total valence electrons. |
| 42 | MaxR_Ele_in_Solid | 0.046 | Maximum range of electrons in solid elements for electron energy of 15 keV (in mm) [43]. |

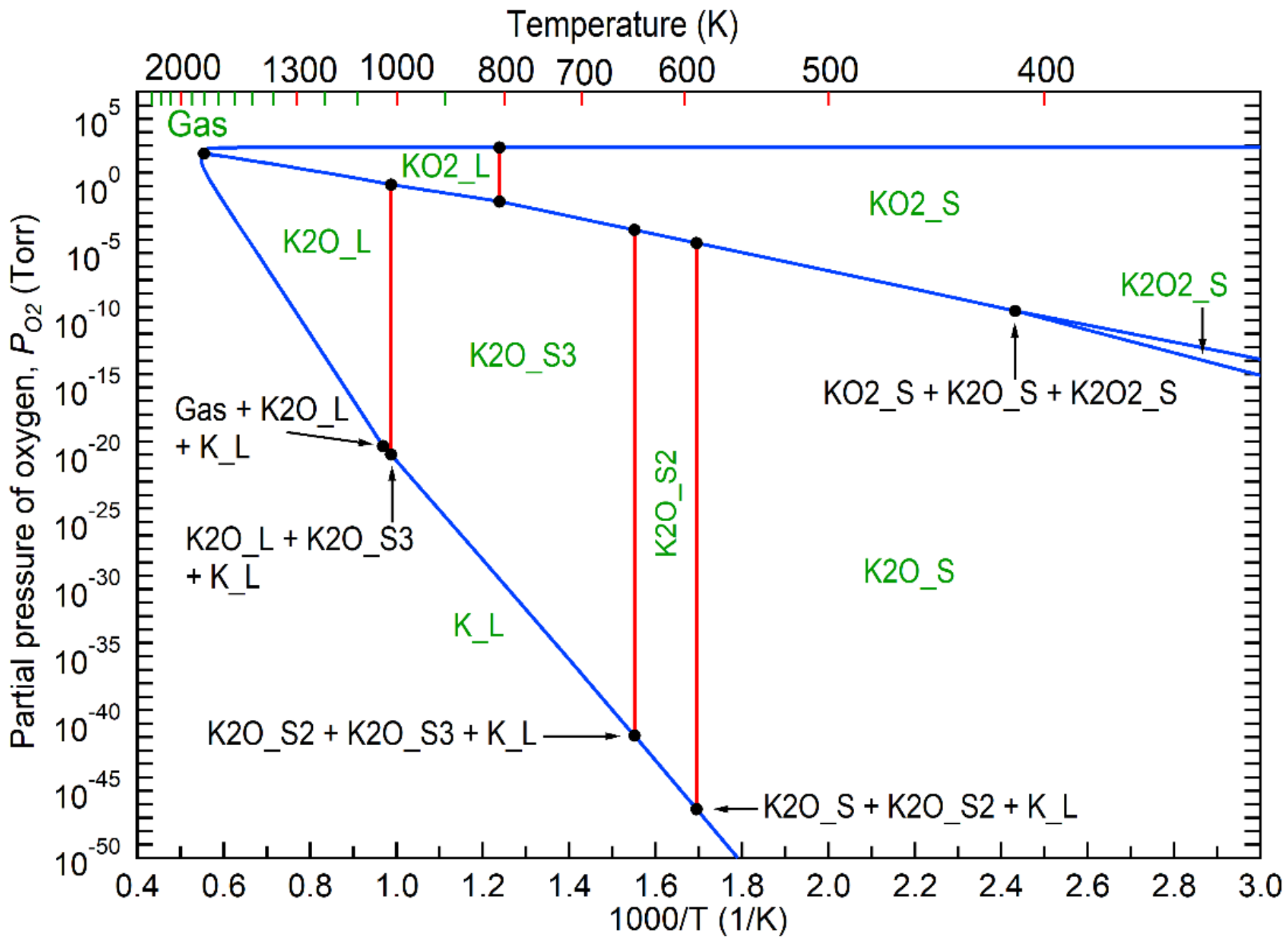

Figure S 1. Calculated K-O pressure-temperature phase diagram in terms of the SSUB5 database (K2O2_S is stable below 411 K). With increasing temperature, the phase transitions follow K → K2O → K2O2 (stable below 411 K) → KO2 → Gas (O2).

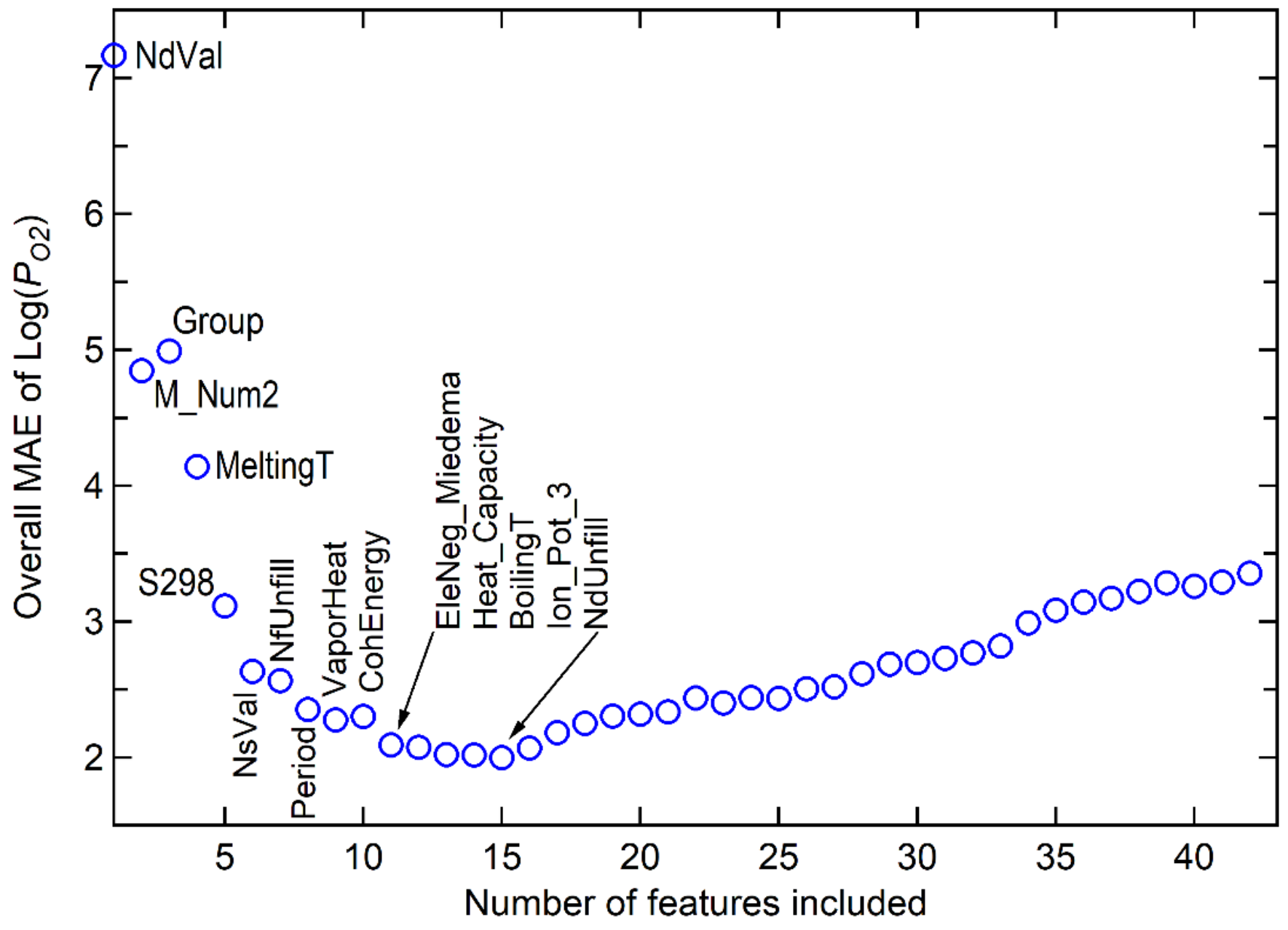

Figure S 2. Examination of the combined features using the sequential feature selection method and the algorithm of Gaussian process regression (with the kernel function of matern52). Among the 42 features, the top 15 features are the present selection as marked in this figure.

**References for both the main and supplemental texts:**

[1] F.H. Stott, G.C. Wood, J. Stringer, The influence of alloying elements on the development and maintenance of protective scales, Oxid. Met. 44 (1995) 113–145. https://doi.org/10.1007/BF01046725.

[2] L. Guo, S.-L. Shang, N. Campbell, P.G. Evans, M. Rzchowski, Z.-K. Liu, C.-B. Eom, Searching for a route to synthesize in situ epitaxial Pr2Ir2O7 thin films with thermodynamic methods, Npj Comput. Mater. 7 (2021) 144. https://doi.org/10.1038/s41524-021-00610-9.

[3] B. Gorr, S. Schellert, F. Müller, H.-J. Christ, A. Kauffmann, M. Heilmaier, Current Status of Research on the Oxidation Behavior of Refractory High Entropy Alloys, Adv. Eng. Mater. 23 (2021) 2001047. https://doi.org/10.1002/adem.202001047.

[4] J. Sun, M.R. Barone, C.S. Chang, M.E. Holtz, H. Paik, J. Schubert, D.A. Muller, D.G. Schlom, Growth of PdCoO 2 by ozone-assisted molecular-beam epitaxy, APL Mater. 7 (2019) 121112. https://doi.org/10.1063/1.5130627.

[5] C.-L. Jia, K.W. Urban, M. Alexe, D. Hesse, I. Vrejoiu, Direct Observation of Continuous Electric Dipole Rotation in Flux-Closure Domains in Ferroelectric Pb(Zr,Ti)O 3, Science (80-. ). 331 (2011) 1420–1423. https://doi.org/10.1126/science.1200605.

[6] G.H. Jonker, J.H. Van Santen, Ferromagnetic compounds of manganese with perovskite structure, Physica. 16 (1950) 337–349. https://doi.org/10.1016/0031-8914(50)90033-4.

[7] A.J. Hauser, R.E.A. Williams, R.A. Ricciardo, A. Genc, M. Dixit, J.M. Lucy, P.M. Woodward, H.L. Fraser, F. Yang, Unlocking the potential of half-metallic Sr2FeMoO6 films through controlled stoichiometry and double-perovskite ordering, Phys. Rev. B. 83 (2011) 014407. https://doi.org/10.1103/PhysRevB.83.014407.

[8] J. Wang, J.B. Neaton, H. Zheng, V. Nagarajan, S.B. Ogale, B. Liu, D. Viehland, V. Vaithyanathan, D.G. Schlom, U. V. Waghmare, N.A. Spaldin, K.M. Rabe, M. Wuttig, R. Ramesh, Epitaxial BiFeO3 Multiferroic Thin Film Heterostructures, Science (80-. ). 299 (2003) 1719–1722. https://doi.org/10.1126/science.1080615.

[9] J.H. Lee, X. Ke, R. Misra, J.F. Ihlefeld, X.S. Xu, Z.G. Mei, T. Heeg, M. Roeckerath, J. Schubert, Z.K. Liu, J.L. Musfeldt, P. Schiffer, D.G. Schlom, Adsorption-controlled growth of BiMnO3 films by molecular-beam epitaxy, Appl. Phys. Lett. 96 (2010) 262905. https://doi.org/10.1063/1.3457786.

[10] A. Schilling, M. Cantoni, J.D. Guo, H.R. Ott, Superconductivity above 130 K in the Hg–Ba–Ca–Cu–O system, Nature. 363 (1993) 56–58. https://doi.org/10.1038/363056a0.

[11] Y. Ma, A. Edgeton, H. Paik, B.D. Faeth, C.T. Parzyck, B. Pamuk, S. Shang, Z.-K. Liu, K.M. Shen, D.G. Schlom, C. Eom, Realization of Epitaxial Thin Films of the Topological Crystalline Insulator Sr3SnO, Adv. Mater. 32 (2020) 2000809. https://doi.org/10.1002/adma.202000809.

[12] H.J.T. Ellingham, Transactions and Communications, J. Soc. Chem. Ind. 63 (1944) 125–160. https://doi.org/10.1002/jctb.5000630501.

[13] H. Paik, Z. Chen, E. Lochocki, A. Seidner H., A. Verma, N. Tanen, J. Park, M. Uchida, S. Shang, B.-C. Zhou, M. Brützam, R. Uecker, Z.-K. Liu, D. Jena, K.M. Shen, D.A. Muller, D.G. Schlom, Adsorption-controlled growth of La-doped BaSnO3 by molecular-beam epitaxy, APL Mater. 5 (2017) 116107. https://doi.org/10.1063/1.5001839.

[14] J.F. Ihlefeld, W. Tian, Z.K. Liu, W.A. Doolittle, M. Bernhagen, P. Reiche, R. Uecker, R.

Rramesh, D.G. Schlom, Adsorption-controlled growth of BiFeO3 by MBE and integration with wide band, in: IEEE Trans. Ultrason. Ferroelectr. Freq. Control, 2009: pp. 1528–1533. https://doi.org/10.1109/TUFFC.2009.1216.

[15] C.M. Brooks, R. Misra, J.A. Mundy, L.A. Zhang, B.S. Holinsworth, K.R. O'Neal, T. Heeg, W. Zander, J. Schubert, J.L. Musfeldt, Z.K. Liu, D.A. Muller, P. Schiffer, D.G. Schlom, The adsorption-controlled growth of LuFe2O4 by molecular-beam epitaxy, Appl. Phys. Lett. 101 (2012) 132907. https://doi.org/10.1063/1.4755765.

[16] E.H. Smith, J.F. Ihlefeld, C.A. Heikes, H. Paik, Y. Nie, C. Adamo, T. Heeg, Z.K. Liu, D.G. Schlom, Exploiting kinetics and thermodynamics to grow phase-pure complex oxides by molecular-beam epitaxy under continuous codeposition, Phys. Rev. Mater. 1 (2017) 023403. https://doi.org/10.1103/PhysRevMaterials.1.023403.

[17] H.P. Nair, Y. Liu, J.P. Ruf, N.J. Schreiber, S.-L. Shang, D.J. Baek, B.H. Goodge, L.F. Kourkoutis, Z.-K. Liu, K.M. Shen, D.G. Schlom, Synthesis science of SrRuO3 and CaRuO3 epitaxial films with high residual resistivity ratios, APL Mater. 6 (2018) 046101. https://doi.org/10.1063/1.5023477.

[18] M.. Sainz, A.. Mazzoni, E.. Aglietti, A. Caballero, Thermochemical stability of spinel (MgO·Al2O3) under strong reducing conditions, Mater. Chem. Phys. 86 (2004) 399–408. https://doi.org/10.1016/j.matchemphys.2004.04.007.

[19] E. Niwa, K. Wakai, T. Hori, K. Yashiro, J. Mizusaki, T. Hashimoto, Thermodynamic analyses of structural phase transition of Pr2NiO4+δ involving variation of oxygen content, Thermochim. Acta. 575 (2014) 129–134. https://doi.org/10.1016/j.tca.2013.10.025.

[20] K. Ishizaki, Phase diagrams under high total gas pressures — Ellingham diagrams for hot isostatic press processes, Acta Metall. Mater. 38 (1990) 2059–2066. https://doi.org/10.1016/0956-7151(90)90073-P.

[21] W. Smeltzer, D. Young, Oxidation properties of transition metals, Prog. Solid State Chem. 10 (1975) 17–54. https://doi.org/10.1016/0079-6786(75)90003-5.

[22] W.M. Mellor, K. Kaufmann, O.F. Dippo, S.D. Figueroa, G.D. Schrader, K.S. Vecchio, Development of ultrahigh-entropy ceramics with tailored oxidation behavior, J. Eur. Ceram. Soc. 41 (2021) 5791–5800. https://doi.org/10.1016/J.JEURCERAMSOC.2021.05.010.

[23] L. Backman, E.J. Opila, Thermodynamic assessment of the group IV, V and VI oxides for the design of oxidation resistant multi-principal component materials, J. Eur. Ceram. Soc. 39 (2019) 1796–1802. https://doi.org/10.1016/j.jeurceramsoc.2018.11.004.

[24] I. Roy, P.K. Ray, G. Balasubramanian, Modeling Oxidation of AlCoCrFeNi High-Entropy Alloy Using Stochastic Cellular Automata, Entropy. 24 (2022) 1263. https://doi.org/10.3390/e24091263.

[25] C.W. Bale, E. Bélisle, P. Chartrand, S.A. Decterov, G. Eriksson, K. Hack, I.-H. Jung, Y.-B. Kang, J. Melançon, A.D. Pelton, C. Robelin, S. Petersen, FactSage thermochemical software and databases — recent developments, Calphad. 33 (2009) 295–311. https://doi.org/10.1016/j.calphad.2008.09.009.

[26] N. Birks, G.H. Meier, F.S. Pettit, Introduction to the High Temperature Oxidation of Metals, 2nd ed., Cambridge University Press, 2006. https://doi.org/10.1017/CBO9781139163903.

[27] B. Gleeson, Thermodynamics and Theory of External and Internal Oxidation of Alloys,

in: Shreir's Corros., Elsevier, 2010: pp. 180–194. https://doi.org/10.1016/B978-044452787-5.00012-3.
[28] Z.-K. Liu, First-principles calculations and CALPHAD modeling of thermodynamics, J. Phase Equilibria Diffus. 30 (2009) 517–534. https://doi.org/10.1007/s11669-009-9570-6.
[29] K.M. Adkison, S.-L. Shang, B.J. Bocklund, D. Klimm, D.G. Schlom, Z.-K. Liu, Suitability of binary oxides for molecular-beam epitaxy source materials: A comprehensive thermodynamic analysis, APL Mater. 8 (2020) 081110. https://doi.org/10.1063/5.0013159.
[30] J.D.A. Simoni, A.P. Chagas, Diagramas de Ellingham e de Van't Hoff: Algumas considerações, Quim. Nova. 30 (2007) 501–504. https://doi.org/10.1590/S0100-40422007000200047.
[31] J.-O. Andersson, T. Helander, L. Höglund, P. Shi, B. Sundman, Thermo-Calc & DICTRA: Computational tools for materials science, Calphad. 26 (2002) 273–312. https://doi.org/10.1016/S0364-5916(02)00037-8.
[32] Scientific Group Thermodata Europe (SGTE), Thermodynamic Properties of Inorganic Materials, in: Lehrstuhl fuer Theoretische Huettenkunde (Ed.), Landolt-Boernstein New Ser. Gr. IV, Springer, Verlag Berlin Heidelberg, 1999.
[33] F. Tesfaye, D. Sukhomlinov, D. Lindberg, P. Taskinen, G. Akdogan, Thermal stabilities and properties of equilibrium phases in the Pt-Te-O system, J. Chem. Thermodyn. 106 (2017) 47–58. https://doi.org/10.1016/j.jct.2016.11.016.
[34] X. Chong, S.-L. Shang, A.M. Krajewski, J.D. Shimanek, W. Du, Y. Wang, J. Feng, D. Shin, A.M. Beese, Z.-K. Liu, Correlation analysis of materials properties by machine learning: illustrated with stacking fault energy from first-principles calculations in dilute fcc-based alloys, J. Phys. Condens. Matter. 33 (2021) 295702. https://doi.org/10.1088/1361-648X/ac0195.
[35] C.E. Rasmussen, C.K.I. Williams, Gaussian processes for machine learning, MIT Press, 2006.
[36] B.B. Newcomb, W.M. Stobbs, E. Metcalfe, A microstructural study of the oxidation of Fe-Ni-Cr alloys I. protective oxide growth, Philos. Trans. R. Soc. London. Ser. A, Math. Phys. Sci. 319 (1986) 191–218. https://doi.org/10.1098/rsta.1986.0097.
[37] A. Ross, S. Shang, H. Fang, G. Lindwall, X.L. Liu, W. Zhao, B. Gleeson, M.C. Gao, Z. Liu, Tailoring critical Al concentration to form external Al2O3 scale on Ni–Al alloys by computational approach, J. Am. Ceram. Soc. 105 (2022) 7770–7777. https://doi.org/10.1111/jace.18707.
[38] A. Zunger, Systematization of the stable crystal structure of all AB-type binary compounds: A pseudopotential orbital-radii approach, Phys. Rev. B. 22 (1980) 5839–5872. https://doi.org/10.1103/PhysRevB.22.5839.
[39] C. Kittel, Introduction to Solid State Physics, John Wiley & Sons, Inc., Hoboken, NJ, 2005.
[40] A.T. Dinsdale, SGTE data for pure elements, Calphad. 15 (1991) 317–425. https://doi.org/10.1016/0364-5916(91)90030-N.
[41] Wolfram Mathematica: Modern Technical Computing, (n.d.). https://www.wolfram.com/mathematica (accessed March 16, 2020).
[42] R.F. Zhang, S.H. Zhang, Z.J. He, J. Jing, S.H. Sheng, Miedema Calculator: A thermodynamic platform for predicting formation enthalpies of alloys within framework

of Miedema's Theory, Comput. Phys. Commun. 209 (2016) 58–69. https://doi.org/10.1016/J.CPC.2016.08.013.
[43] G. V. Samsonov, Handbook of the Physicochemical Properties of the Elements, Springer, New York, 1968. https://doi.org/10.1007/978-1-4684-6066-7_1.
[44] The Periodic Table, (n.d.). https://periodictable.com (accessed March 16, 2020).
[45] H.B. Michaelson, The work function of the elements and its periodicity, J. Appl. Phys. 48 (1977) 4729–4733. https://doi.org/10.1063/1.323539.
[46] Periodic Table of Elements - PubChem, (n.d.). https://pubchem.ncbi.nlm.nih.gov/periodic-table/ (accessed March 16, 2020).
[47] Elastic properties of the elements (data page) - Wikipedia, (n.d.). https://en.wikipedia.org/wiki/Elastic_properties_of_the_elements_(data_page) (accessed March 16, 2020).
[48] AZOM Materials, (n.d.). https://www.azom.com/ (accessed March 16, 2020).
[49] M.S. Anderson, C.A. Swenson, D.T. Peterson, Experimental equations of state for calcium, strontium, and barium metals to 20 kbar from 4 to 295 K, Phys. Rev. B. 41 (1990) 3329–3338. https://doi.org/10.1103/PhysRevB.41.3329.